\documentclass[prd,twocolumn,amsmath,showpacs,amssymb,floatfix,superscriptaddress,nofootinbib]{revtex4}

\usepackage{amsfonts,color,xcolor,graphicx,amsfonts,multirow,amsmath}
\usepackage[colorlinks=true,citecolor=blue,linkcolor=blue,urlcolor=blue]{hyperref}
\usepackage{setspace}
\allowdisplaybreaks
\newcommand{\ad}{\hspace{-0.3cm}}
\begin{document}

\title{Investigating $D_s^+ \to \pi^0 \ell^+ \nu_\ell$ decay process within QCD sum rule approach}
\author{Hai-Jiang Tian}
\author{Hai-Bing Fu}
\email{fuhb@gzmu.edu.cn (corresponding author)}
\author{Tao Zhong}
\email{zhongtao1219@sina.com}
\address{Department of Physics, Guizhou Minzu University, Guiyang 550025, P.R.China}
\author{Xuan Luo}
\address{School of Physics, Southeast University, Nanjing 210094, China}
\author{Dan-Dan Hu}
\address{Department of Physics \& Chongqing Key Laboratory for Strongly Coupled Physics, Chongqing University, Chongqing 401331, P.R. China}
\author{Yin-Long Yang}
\address{Department of Physics, Guizhou Minzu University, Guiyang 550025, P.R.China}

\begin{abstract}
In this paper, the semileptonic decays $D_s^+ \to \pi^0\ell^+ \nu_\ell$ with $\ell=(e,\mu)$ are investigated by using the light-cone sum rule approach. Firstly, the neutral meson mixing scheme between $\pi^0$, $\eta$, $\eta^\prime$ and pseudoscalar gluonium $G$ is discussed in a unified way, which leads to the direct connection between two different channels for $D_s^+\to \pi^0\ell^+\nu_\ell$ and $D_s^+ \to \eta\ell^+\nu_\ell$ by the $\pi^0-\eta$ mixing angle. Then we calculated the $D_s\to \pi^0$ transition form factors (TFFs) within QCD light-cone sum rule approach up to next-to-leading order correction. At the large recoil point, we have $f_+^{D_s^+\pi^0}(0)=0.0113_{-0.0019}^{+0.0024}$ and $f_-^{D_s^+\pi^0}(0)=0.0020_{-0.0009}^{+0.0008}$. Furthermore, the TFFs are extrapolated to the whole physical $q^2$-region by using the simplified $z(q^2)$-series expansion. The behaviors of TFFs and related three angular coefficient functions $a_{\theta_\ell}(q^2)$, $b_{\theta_\ell}(q^2)$ and $c_{\theta_\ell}(q^2)$ are given. The differential decay widths for $D_s^+ \to \pi^0\ell^+ \nu_\ell$ with respect to $q^2$ and $\cos\theta_\ell$ are presented, and also lead to the branching fractions ${\cal B}(D_s^+\to \pi ^0e^+\nu_e) =2.60_{-0.51}^{+0.57}\times 10^{-5}$ and ${\cal B}(D_s^+\to \pi ^0\mu^+\nu _\mu )= 2.58_{-0.51}^{+0.56}\times 10^{-5}$. These results show well agreement with the recent BESIII measurements and theoretical predictions. Then the differential distributions and integrated predictions for three angular observables, {\it i.e.} forward-backward asymmetries, $q^2$-differential flat terms and lepton polarization asymmetries are given separately. Lastly, we estimate the ratio for different decay channels ${\cal R}_{\pi ^0/\eta}^{\ell}=1.108_{-0.071}^{+0.039}\times 10^{-3}$.
\end{abstract}

\date{\today}


\maketitle
\section{Introduction}
Since the development of QCD reveals the observed mixing pattern of isospin mesons, the meson mixing effect is recognized as one of the topics of considerable interest, which can provide an explanation for the disparity between valence states of $I=0$ pseudoscalar and vector mesons~\cite{Khodjamirian:2020btr}. There are two schemes that frequently adopted by researchers in dealing with meson mixing: the octet-singlet mixing scheme and the quark-flavor mixing scheme. These two schemes have been extensively investigated both from experimental side~\cite{CELLO:1990klc, TPCTwoGamma:1990dho, KLOE:2002jed, Muller:2004vf, BaBar:2006ash, KLOE:2006guu, Anisovich:1997dz} and theoretical side~\cite{Hu:2021zmy,Huang:2006as,Feldmann:2002kz,Kroll:2004rs, Ball:1995zv,Feldmann:1998su,Feldmann:1998vh,Feldmann:1998sh,Tippens:2001fq}. In order to further understand dynamics and hadronic structure, mixing among pseudoscalar mesons will lead to QCD anomaly and is connected with chiral symmetry breaking. Without a doubt, one could understand the dynamics more clearly if the mixing parameters were specified with better fidelity. On the other hand, when neutral mesons have the same quantum numbers and hidden flavors, they will mix with each other through the strong and electromagnetic interactions, which can also be used to explain some particular heavy meson decay processes. Such as the systems $\pi^0-\eta$~\cite{Li:2020ylu,BESIII:2022jcm}, $\eta-\eta^\prime$~\cite{Benayoun:1999fv, Ricciardi:2012xu, Ke:2010htz, Choi:2010zb}, $\omega-\phi$~\cite{Gronau:2009mp,Kucukarslan:2006wk,Gronau:2008kk}, and $\rho-\omega$~\cite{Maltman:1995nq, Maltman:1996kj, OConnell:1997ggd,Gardner:1997yx}.

\begin{figure*}[t]
\centering
\includegraphics[width=1\textwidth]{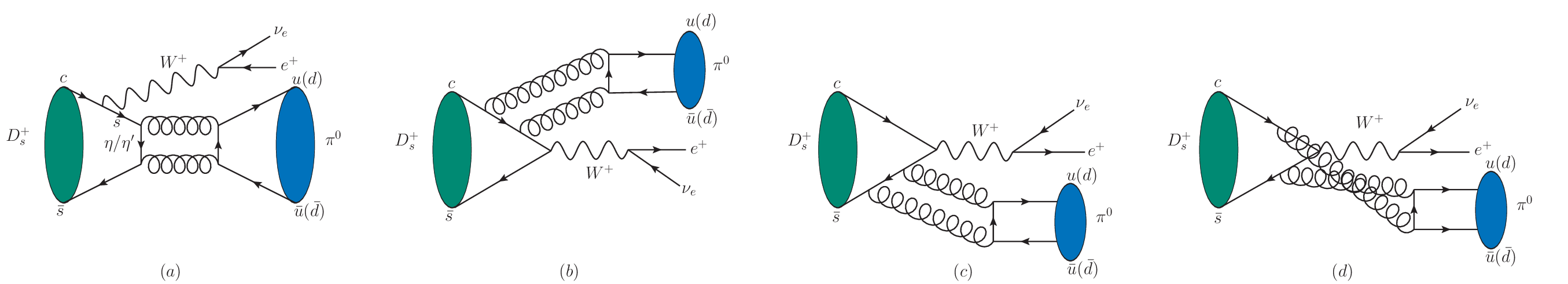}
\caption{Diagrams for the semileptonic decay $D_s^+\to\pi^0\ell^+\nu_\ell$ indicate that this process can only occur through $\pi^0-\eta$ mixing indicated by subdiagram $(a)$ and nonperturbative weak annihilation effects with the radiation of a $\pi^0$-meson represented by $(b)$, $(c)$, $(d)$, respectively.}
\label{Fig:1}
\end{figure*}

Recently in 2022, the BESIII collaboration reported its first observation for $D_s^+ \to \pi^0 e^+ \nu_e$ decay process and reported measured upper limits of branching fraction, {\it i.e.} $6.4\times10^{-5}$ by using a data sample of electron-positron collisions corresponding to an integrated luminosity of 6.32 fb$^{-1}$ at center-of-mass energies between 4.178 and 4.226 GeV~\cite{BESIII:2022jcm}. The $D_s^+ \to \pi^0e^+ \nu_e$ decay process can be investigated by the using neutral meson mixing scheme, which resembles the $D_s^+\to \omega e^+\nu_e$ process. This process can be occurred via $\pi^0-\eta$ meson mixing and nonperturbative weak annihilation (WA), which can provide an excellent platform for studying the meson mixing effect and its associated preference~\cite{Gronau:2009mp}. The $D_s^+$ meson composed of $c\bar{s}$ system decays into $\pi^0$ meson, which is expected to relate to the small admixture of $s\bar{s}$ in the wave function of the $\pi^0$ meson that originates from the mixing of $\pi^0-\eta$ meson. Due to Okubo-Zweig-Iizuka rule (OZI)~\cite{Okubo:1963fa, Zweig:1964ruk, Iizuka:1966fk} and isospin violation~\cite{Li:2020ylu}, the WA effect in the $D_s^+ \to \pi^0e^+ \nu_e$ decay process is suppressed and its value is only have $10^{-7}-10^{-8}$ order~\cite{BESIII:2022jcm}. Meanwhile, $\pi^0-\eta$ meson mixing to the process $D_s^+ \to \pi^0e^+ \nu_e$ will reach to $10^{-5}$ order, which is significantly different with WA effective. Thus, one can investgate the $\pi^0-\eta$ mixing effect more accuracy rather than WA effect. According to the scheme of neutral meson mixing, the relationship between the $D_s^+\to \pi$, $D_s^+ \to \eta$ transition form factors (TFFs) and mixing angle $\delta$ can be adapted. So the full analytical expression for the TFFs should be taking into consideration. Moreover, there are some early studies on the flavor-symmetry breaking of $D\to P\ell\nu$ pseudoscalar meson decay~\cite{Khlopov:1978id, Gershtein:1976mv}. And the study of the semilptonic decays of heavy flavor mesons not only offers a clean environment to extract the CKM matrix element but also describes the CP-violating and flavor-changing processes in the SM, which have an explicit statement and investigation in Refs~\cite{Zhang:2020dla, Faustov:2019mqr}, providing a good motivation for our subsequent calculation.

At present, there are various approaches to study TFFs, such as the lattice QCD (LQCD)~\cite{Bali:2014pva}, traditional or covariant light-front quark model (LFQM)~\cite{Cheng:2017pcq,Verma:2011yw,Wei:2009nc}, constituent quark model (CQM)~\cite{Melikhov:2000yu}, covariant confined quark model (CCQM)~\cite{Soni:2018adu,Ivanov:2019nqd}, QCD sum rules (QCDSR)~\cite{Colangelo:2001cv} and light-cone sum rules (LCSR)~\cite{Offen:2013nma,Duplancic:2015zna}. Among these approaches, the LCSR affords an efficient method in making predictions for exclusive processes, which allows incorporating information about high-energy asymptotic correlation functions in QCD that change into light-cone distribution amplitudes (LCDAs). So the important component in TFFs is the meson's LCDAs, which are related to the nonlocal light-ray operators between the hadronic state and vacuum. In this paper, $\eta$-meson twist-2 LCDA that takes main contribution is calculated by using the QCDSR approach under background field theory.

The rest of the paper are organized as follows: In Sec.~\ref{Sec:2}, we present the basic idea for the neutral mesons mixing mechanism and the decays $D_s^+\to\pi^0 \ell^+\nu_\ell$, and also give TFFs for the transition $D_s^+\to\pi^0$. In Sec.~\ref{Sec:3}, we present the numerical analysis. Section~\ref{Sec:4} is a brief summary.

\section{Theoretical framework}\label{Sec:2}
The decay process $D_s^+ \to \pi^0 \ell^+ \nu_\ell$ can be represented by the four typical diagrams, which are shown in Fig.~\ref{Fig:1}. The panel (a) is the $\pi^0-\eta$ mixing and the panels (b)-(c) are the nonperturbative WA effects with the radiation of a $\pi^0$ meson, respectively. In panel (a), there is a small component of $s\bar{s}$ in the wave function of $\pi^0$ due to the mixing of $\pi^0-\eta$ mesons. Therefore the transition of $D_s^+$ can be induced by the $s\bar{s}$ component in $\pi^0$ via $D_s^+\to(s\bar{s})e^+\nu_e$ transition. Meanwhile, the $c$ quark emitted a $W^+$ boson and $s$ quark in which the former changes into two leptons and the latter together with $\bar{s}$ quark constitute a small $s\bar{s}$ component that can be regarded as existing in the wave function of $\pi^0$. Subsequently, the bound state is formed by the emission of the two gluons~\cite{Li:2020ylu, BESIII:2022jcm, Gronau:2009mp}. The panel (b) represents the two gluons are emitted from a $c$ quark; the panel (c) stands for two gluons are emitted from a $\bar{s}$ quark; the panel (d) refers to one gluon is emitted from each quark. The corresponding descriptions can also be found in reference from BESIII collaboration recently in 2022~\cite{BESIII:2022jcm}. All three processes are caused by the nonperturbative weak annihilation effect.  Especially, the panel (c)  is one of the three various processes caused by the nonperturbative weak annihilation effect, which occurs by the preradiation of a $\pi^0$ meson from the $c\bar{s}$ system in $D_s^+$ meson, followed by the weak transition of $c\bar{s}\to e^+\nu_e$~\cite{Li:2020ylu}. Meanwhile, there is no $s\bar{s}$ component in panel (c), which is distinguished with the process of panel (a). Moreover, the weak annihilation involves OZI-suppressed nonperturbative preradiation of an isoscalar system such as $\omega$ meson, {\it e.g.} in $D_s \to \omega(D_s^*)_{\mathrm{virtual}}\to\omega\ell\nu$~\cite{Gronau:2009mp}.

The neutral meson mixing one is dominant, and its particular property is to mix the $\pi^0, \eta, \eta^\prime, G$ in each pair by a unified way~\cite{Li:2020ylu}. Four physical states are taken to be linear combinations with these flavor bases, {\it i.e}.
\begin{eqnarray}\label{Eq::1}
\left[ \begin{array}{c}
	\pi ^0\\
	\eta\\
	\eta '\\
	\eta _G\\
\end{array} \right] =V\left[ \begin{array}{c}
	\pi_q^0\\
	\eta_q\\
	\eta_s\\
	G\\
\end{array} \right],
\end{eqnarray}
where $\pi_q^0=(u\bar u-d\bar d)/\sqrt2$, $\eta_q = (u\bar u+d\bar d)/\sqrt2$, $\eta_s = s\bar{s}$ and $G$ as the pure pseudoscalar gluonium. Meanwhile, the $(4\times 4)$ real matrix $V$ should have 6 independent parameters to keep the unitary and can be regarded as the mixing angles between these mesons. Due to the fact that angles for $\pi^0_q-\eta_s$ and $\pi^0_q-G$ mixing are isospin-violating, there exists a large mass gap among them. This will lead to mixing angles tending to zero approximately. Meanwhile, the $\eta_q-G$ mixing angle also tends to zero, where detailed analysis are shown in Ref.~\cite{DeFazio:2000my}. For the other three mixing angles, we can explicitly write the sub-mixing matrix as follows~\cite{Li:2020ylu}:
\begin{eqnarray}
V_1(\pi^0-\eta_q) \!\!&=&\!\!
\left[ \begin{matrix}
+\cos\delta&	\,\,-\sin\delta~&	~~~~0~~~~&		~~~~~0~~~~~\\
+\sin\delta&	+\cos\delta&		0&		0\\
	0&		0&		1&		0\\
	0&		0&		0&		1\\
\end{matrix} \right],\\
V_2( \eta_q-\eta_s ) \!\!\!&=&\!\!
\left[ \begin{matrix}
~~~~1~~~~&		0&		0&		~~~~0~~~~~ \\
~~~~0~~~~&		+\cos\phi&		~-\sin \phi~&		~~~~0~~~~~\\
~~~~0~~~~&		+\sin\phi&		+\cos \phi&		~~~~0~~~~~\\
~~~~0~~~~&		0&		0&		~~~~1~~~~~\\
\end{matrix} \right],\\
V_3( \eta_s-G ) \!\!&=&\!\!
\left[ \begin{matrix}
~~~~1~~~~&		~~~~0~~~~&		0&		0\\
~~~~0~~~~&		1&		0&		0\\
~~~~0~~~~&		0&	+\cos\phi_G  &	+\sin\phi_G\\
~~~~0~~~~&		0&	-\sin \phi_G &	+\cos\phi_G\\
\end{matrix} \right].
\end{eqnarray}
In which, the symbol $\delta$ stands for the mixing angle of $\pi^0_q$ and $\eta_q$, $\phi$ denotes the mixing angle of $\eta_q$ and $\eta_s$, $\phi_G$ represents the mixing angle of $\eta_s$ and $G$.  After combing the above three mixing matrix, {\it i.e.} $V=V_3V_1V_2$ , one can get the following expression
\begin{align}
V\simeq& \left[ \begin{matrix}
1      &		-\delta \cos \phi     &		+\delta \sin \phi     &		0          \\
\delta &		+\cos \phi            &		-\sin \phi            &		0           \\
0      &		+\cos \phi_G\sin \phi &		+\cos \phi_G\cos \phi &		+\sin \phi_G\\
0      &		-\sin \phi_G\sin \phi &		-\sin \phi_G\cos \phi &		+\cos \phi_G\\
\end{matrix} \right].
\label{Eq::5}
\end{align}
The mixing angle $\delta$ is small due to the isospin-violating. Then, we can get the following four equations by comparing with  Eqs.~\eqref{Eq::1} and \eqref{Eq::5}, respectively:
\begin{eqnarray}
&&\ad|\pi ^0\rangle =|\pi_q^0\rangle -\delta \cos \phi |\eta_q\rangle +\delta \sin \phi |\eta_s\rangle,
\label{Eq::6}
\\
&&\ad|\eta \rangle =\delta |\pi_q^0\rangle +\cos \phi |\eta_q\rangle -\sin \phi |\eta_s\rangle,
\label{Eq::7}
\\
&&\ad|\eta'\rangle =\cos \phi_G\sin \phi |\eta_q\rangle+\cos \phi_G\cos \phi |\eta_s\rangle+\sin \phi_G|G\rangle,
\label{Eq::8}
\\
&&\ad|\eta_G\rangle =-\sin \phi_G\sin \phi |\eta_q\rangle -\sin \phi_G\cos \phi |\eta_s\rangle+\cos \phi_G|G\rangle.
\nonumber\\
\label{Eq::9}
\end{eqnarray}

Then, one can obtain the relationships among the transition matrix elements of $\langle\pi^0|V_\mu|D_s^+\rangle$, $\langle\eta|V_\mu| D_s^+ \rangle$, $\langle\eta_s|V_\mu|D_s^+\rangle$ with the help of Eqs. \eqref{Eq::6} and \eqref{Eq::7},
\begin{eqnarray}
&&\langle \pi ^0| V_{\mu} | D_s^+ \rangle  =\delta \sin \phi \langle \eta_s| V_{\mu} | D_s^+ \rangle,
\\
&&\langle \eta | V_{\mu} | D_s^+ \rangle  = - \sin \phi \langle \eta_s| V_{\mu} | D_s^+ \rangle.
\end{eqnarray}
The transitions $D^+_s\to\pi^0$ and $D^+_s\to\eta$ are induced via the component $s\bar{s}$, which can be seen in Fig.~\ref{Fig:1}(a). Due to the transition matrix elements $\langle P| V_{\mu} | D_s^+ \rangle$ with $P = (\pi^0, \eta_s)$ have the definition
\begin{eqnarray}
\ad\langle P (p)|V_\mu|D_s^+(p+q)\rangle =2f_+^{D_s^+P}(q^2) p_{\mu}+\tilde{f}^{D_s^+P}(q^2) q_{\mu},
\end{eqnarray}
with relationship $\tilde{f}^{D_s^+P} (q^2) = f_+^{D_s^+P}(q^2) + f_-^{D_s^+P}(q^2)$ and $q$ being the momentum transfer.
Therefore, one can obtain the relationship between the two TFFs $f^{D_s^+ \pi^0 (\eta)}_{\pm}(q^2)$ and $f^{D_s^+\eta_s}_{\pm}(q^2)$:
\begin{eqnarray}
&&f_\pm^{D_s^+\pi^0}(q^2) = \delta \sin \phi f_\pm^{D_s^+\eta_s}(q^2),
\label{Eq:9}
\\
&&f_\pm^{D_s^+\eta}(q^2) =-\sin \phi f_\pm^{D_s^+\eta_s}(q^2).
\label{Eq:10}
\end{eqnarray}
By comparing Eqs.~(\ref{Eq:9}) and (\ref{Eq:10}), we can acquire the relational expression:
\begin{eqnarray}\label{Eq::14}
\frac{f_\pm^{D_s^+\pi^0}(q^2)}{f_\pm^{D_s^+\eta}(q^2)}=-\delta.
\end{eqnarray}

To calculate the mixing angle $\delta$, there have two schemes. The first one is to expand the mixing angle as lowest-order $\delta^{(2)}$ and higher-order terms $\delta^{(4)}$, which can be expressed as $\delta=\delta^{(2)}+\delta^{(4)}$. The $\delta^{(2)}$ can be expressed in terms of quark mass ratios, and the higher-order term $\delta^{(4)}$ requires another scheme to obtain, whose detailed expression and calculation approach can be seen in the Refs.~\cite{Gasser:1984ux,Ecker:1999kr}. The second one is to use the ratio of $\eta^\prime\to\pi^+\pi^-\pi^0$ and $\eta^\prime\to\pi^+\pi^-\eta$ branching fractions, where the former is $G$-parity violating that can only occur through $\pi^0-\eta$ mixing. Due to second method can be determined from the experimental side and our calculation of this paper connected with the meson mixing scheme directly, so we will take the second scheme. The ratio of the decay branching fractions has the following form:
\begin{align}\label{Eq::18}
\frac{{\cal B}(\eta '\to \pi^+\pi^-\pi ^0 )}{{\cal B}(\eta '\to \pi^+\pi^-\eta )}&=\bigg| \frac{\langle \pi^+\pi^-\pi ^0|H| \eta ' \rangle}{\langle \pi^+\pi^-\eta |H| \eta ' \rangle} \bigg|^2\frac{\phi _s( \eta '\to \pi^+\pi^-\pi ^0 )}{\phi _s( \eta '\to \pi^+\pi^-\eta )}
\nonumber \\
&=\delta ^2\frac{\phi _s( \eta '\to \pi^+\pi^-\pi ^0 )}{\phi _s( \eta '\to \pi^+\pi^-\eta )},
\end{align}
where $\langle \pi^+\pi^-\pi^0(\eta)| H | \eta ' \rangle$ are the decay amplitudes of $\eta^\prime\to\pi^+\pi^-\pi^0$ and $\eta^\prime\to\pi^+\pi^-\eta$ and $H$ is Hamiltonian that induces the $\eta^\prime$ three-body decays. One can obtain this result: ${\langle\pi^+\pi^-\pi^0|H|\eta'\rangle}/{\langle \pi^+\pi^-\eta | H |\eta'\rangle}=-\delta$ according to the mixing scheme given in Eqs. (\ref{Eq::6}) and (\ref{Eq::7}). Furthermore, $\phi_s(\eta^\prime\to\pi^+\pi^-\pi^0(\eta))$ is the phase space volume of the decay model $\eta^\prime\to\pi^+\pi^-\pi^0(\eta)$. For the ratio $\phi_s(\eta^\prime\to\pi^+\pi^-\pi^0)/\phi_s(\eta^\prime\to\pi^+\pi^-\eta)=17.0$, it can be obtained directly from Refs.~\cite{Gross:1979ur, Cheng:2018smm}. The CLEO and BESIII collaborations have measured the branching fraction of $\eta^\prime\to\pi^+\pi^-\pi^0$, respectively. In 2018, the ratio ${\cal B}(\eta '\to \pi^+\pi^-\pi ^0 )/{\cal B}(\eta '\to \pi^+\pi^-\eta )$ was analyzed based on the data from BESIII, and its value is determined to be $(8.8\pm1.2)\times10^{-3}$ as in Ref.~\cite{Fang:2017qgz}. So, we can obtain the value of $\pi^0-\eta$ mixing angle $\delta$:
\begin{eqnarray}
\delta ^2=(5.18\pm 0.71) \times 10^{-4}.
\end{eqnarray}
The other two mixing angles $\phi$ and $\phi_G$ can be calculated by adopting the $\eta-\eta'-G$ mixing scheme, due to the smallest value of $\delta$. Furthermore, one can consider several decay rates by using the Vector-meson Dominance Model and SU$_{f}$(3) quark model~\cite{Kou:1999tt}, which can be expressed in terms of i) the masses of light mesons; ii) the decay constants of light mesons; iii) the mixing angles~\cite{Ke:2011fj, Kou:1999tt}. After taking the newly average value for several decay rates from Particle Data Group (PDG)~\cite{ParticleDataGroup:2022pth}, The numerical value of the two mixing angles are $\cos ^2 \phi_{G}\approx 0.87_{-0.02}^{+0.02}$ and $\phi\approx (41.841^{+0.230}_{-0.223})^{\circ}$, which are comparable with the KLOE collaboration predictions,  {\it i.e.} $\cos ^2 \phi_G=0.86\pm0.04$ and $\phi=(41.4\pm0.3_{\rm stat.}\pm0.7_{\rm syst.}\pm0.6_{\rm th.})^{\circ}$ within uncertainties~\cite{KLOE:2006guu}.

In order to study relevant physical observaboles, we adopt the explict expression for the full differential decay width distribution of $D_s^+\to P\ell ^+\bar{\nu}_\ell$ as follows:
\begin{align}
\frac{d^2\Gamma (D_s^+ \to P\ell ^+\bar{\nu}_\ell )}{dq^2 d\cos \theta_\ell}& = a_{\theta _\ell}(q^2)+b_{\theta_\ell}(q^2) \cos\theta_\ell \nonumber \\
&+c_{\theta_\ell}(q^2) \cos^2\theta_\ell,
\end{align}
where the three $q^2$-dependent angular coefficient functions have the following expressions~\cite{Becirevic:2016hea,Cui:2022zwm}
\begin{align}
a_{\theta_\ell}(q^2)&={\cal N}_{\rm ew} \lambda^{3/2} \bigg(1 - \frac{m_\ell^2}{q^2} \bigg)^2 \bigg[ \big|f^{D_s^+P}_+\big|^2
\nonumber \\
&+\frac{1}{\lambda}\frac{m_\ell^2}{q^2}\bigg(1-\frac{m_{\pi^0}^2}{m_{D_s^+}^2} \bigg)^2 \big|f^{D_s^+P}_0\big|^2\bigg],
\\
b_{\theta_\ell}(q^2)&=2{\cal N}_{\rm ew}\lambda \bigg(1-\frac{m_\ell^2}{q^2}\bigg)^2\frac{m_\ell^2}{q^2} \bigg(1-\frac{m_P^2}{m_{D_s^+}^2} \bigg)^2
\nonumber \\
&\times \mathrm{Re}\bigg[f^{D_s^+P}_+(q^2) f^{D_s^+P*}_0(q^2) \bigg],
\\
c_{\theta_\ell}(q^2)&=-{\cal N}_{\rm ew}\lambda^{3/2}\bigg(1-\frac{m_\ell^2}{q^2}\bigg)^3 \big|f^{D_s^+P}_+\big|^2.
\end{align}
Here, the scalar form factor $f_{0}^{D_s^+P}(q^2)=f_+^{D_s^+P}(q^2) + q^2/(m_{D_s^+}^2-m_{\pi^0}^2)f_-^{D_s^+P}(q^2)$. For convenience, we have introduced the following shorthand notations, ${\cal N}_{\rm ew} = G_F^2|V_{cs}|^2m_{D_s^+}^3/(256\pi^3)$ and
$\lambda \equiv \lambda (1,m_P^2/m_{D_s^+}^2,q^2/m_{D_s^+}^2)$ with
$\lambda (a,b,c) \equiv a^2 + b^2 + c^2 -2(ab+ac+bc)$.
In this paper, the symbol $P$ can be taken as a $\pi^0$ meson and $G_F=1.166 \times 10^{-5}~{\rm GeV^{-2}}$ is the Fermi coupling constant. $|V_{cs}|$ is the Cabibbo-Kobayashi-Maskawa (CKM) matrix element. With the help of resultant three $q^2$-dependent angular coefficient functions, one can calculate the three differential distribution of angle observables of the semileptonic decay $D_s^+\to \pi^0\ell^+\nu_\ell$ for the forward-backward asymmetries, the $q^2$-differential flat terms, and lepton polarization asymmetry, {\it i.e.} $\mathcal{A}_{\mathrm{FB}}^{D_s^+\to\pi^0\ell^+\nu_\ell}(q^2)$, $\mathcal{F}_{\mathrm{H}}^{D_s^+\to\pi^0\ell^+\nu_\ell}(q^2)$ and $\mathcal{A}_{\lambda_\ell}^{D_s^+\to\pi^0\ell^+\nu_\ell}(q^2)$, respectively. The three observables are extremely sensitive to the leptonic mass and effects of physics beyond the standard model, while a subset of these observables also seemed to be sensitive to the hadrnoic uncertainties~\cite{Becirevic:2016hea}. The detailed expressions can be found in Ref.~\cite{Becirevic:2016hea,Cui:2022zwm}.

Nextly, to derive the $D_s^+\to\pi^0$ TFFs, one can use the QCD LCSR approach. After considering the relationship between TFFs from different channels, {\it i.e.} Eq.~\eqref{Eq::14}, we will take the following correlation function to derive the $f_\pm^{D_s^+\pi^0}(q^2)$ \cite{Descotes-Genon:2019bud}:
\begin{eqnarray}
\Pi_\mu(p,q)= -i\delta \int{d^4e^{iq\cdot x}}\langle\eta(p)|{\rm T}\{j_\mu(x),j_5^\dag(0)\}|0\rangle,
\end{eqnarray}
where $j_\mu(x)=\bar s(x)\gamma_\mu c(x)$, $j_5(x)=m_c\bar s(x)i\gamma _5c(x)$. In the time-like $q^2$-region, we can insert the complete intermediate states that have the same quantum numbers as the current operator $(\bar c i\gamma_5s)$ into the hadron current of the correlation function. After isolating the pole term of the lowest pseudoscalar $D_s$-meson, we can reach the hadronic representation. The dispersion integrations can be replaced with the sum of higher resonances and continuum states. Meanwhile, we work in the space-like $q^2$-region, where the $c$-quark operator needs to contract by applying a propagator with the gluon field correction~\cite{Hu:2021zmy}. For the desired sum rule for TFFs, we need to use the OPE method by considering the meson LCDAs~\cite{Ball:2006wn, Fu:2020uzy}. After using the Borel transformation and substracting the contribution from higher  resonances and continuum states, the LCSR for TFFs can be achieved, it finally read off:
\begin{align}
&f_+^{D_s^+\pi^0}(q^2) = -\delta \frac{e^{m_{D_s^+}^2/M^2}}{2m_{D_s^+}^2f_{D_s^+}}
\nonumber\\
&\qquad \times \bigg[F_0(q^2,M^2,s_0)+ \frac{\alpha_ s C_F}{4\pi}F_1(q^2,M^2,s_0)\bigg],
\\
&\tilde f^{D_s^+\pi^0}(q^2) = -\delta  \frac{e^{m_{D_s^+}^2/M^2}}{m_{D_s^+}^2f_{D_s^+}}
\nonumber\\
&\qquad \times \bigg[\tilde{F}_0(q^2, M^2, s_0) + \frac{\alpha_ s C_F}{4\pi}\tilde{F}_1(q^2, M^2, s_0) \bigg].
\end{align}
The leading-order and next-to-leading order invariant amplitudes $F_0(q^2, M^2, s_0)/\tilde{F}_0(q^2, M^2, s_0)$ and $F_{1}(q^2,M^2,s_0)/\tilde{F}_1(q^2,M^2,s_0)$ are given in Ref.~\cite{Duplancic:2008ix}. The specific detailed expressions are consistent with literature~\cite{Duplancic:2015zna}, which is also discussed in our previous work~\cite{Hu:2021zmy}

\section{Numerical results and discussions}\label{Sec:3}
Before proceeding further calculation, the following choice of input parameters are required. The charm-quark mass is $m_c = 1.27 \pm 0.02$~GeV, $s$-quark mass $m_s = 0.093$ GeV, and the masses of $D_s$, $\eta$, $\pi^0$-meson $m_{D_s} = 1.9685$ GeV, $m_\eta = 0.5478$ GeV, $m_{\pi^0} = 0.13498$ GeV. All of them are taken from the Particle Data Group (PDG)~\cite{ParticleDataGroup:2020ssz}. The $D_s$, $\eta$-meson decay constants are taken as $f_{D_s} = 0.274 \pm 0.013 \pm 0.007$ GeV~\cite{Azizi:2010zj}, $f_\eta = 0.130 \pm 0.003$ GeV~\cite{Ball:2004ye}.

Furthermore, the twist-2, 3, 4 LCDAs for $\eta$-meson are needed. For the twist-2 LCDAs $\phi_{2;\eta}(x,\mu)$, we calculated its first three $\xi$-moments $\langle \xi _{2;\eta}^{n} \rangle |_{\mu}$ with $n=(2,4,6)$ by using the QCD sum rule within background field theory, where the accuracy is up to dimension-six nonperturbative vacuum condensates and next-to-leading QCD correction for the perturbative part. The values are
\begin{align}
&\langle \xi _{2;\eta}^2 \rangle |_{\mu_k}=0.231^{+0.010}_{-0.013},
\nonumber\\
&\langle \xi _{2;\eta}^4 \rangle |_{\mu_k}=0.109^{+0.007}_{-0.007},
\nonumber\\
&\langle \xi _{2;\eta}^6 \rangle |_{\mu_k}=0.066^{+0.006}_{-0.006},
\end{align}
where the typical scale in this paper is taken as $\mu_k = (m_{D_s^+}^2 - m_c^2)^{1/2} \approx 1.5~{\rm GeV}$. Thus, we can obtain higher-order Gergenbauer moments: $a^2_{2;\eta}(\mu_k)=0.089_{0.035}^{+0.030}$, $a^4_{2;\eta}(\mu_k)=0.025_{-0.010}^{+0.003}$, $a^6_{2;\eta}(\mu_k)=0.033_{-0.049}^{+0.054}$. The detailed analysis and calculation processes for $\langle \xi _{2;\eta}^{n} \rangle |_{\mu}$ are shown in our recent work~\cite{Hu:2021zmy}. The twist-3 and twist-4 LCDAs expressions and corresponding parameters are mainly taken from Refs.~\cite{Huang:2001xb, Ball:2006wn}. One could run those hadronic parameters of the twist-2,3,4 LCDAs from the initial factorization scale to other scale, which also requires using the renormalization group equation,
\begin{align}
c_i(\mu_k) =\mathcal{L}^{\gamma_{c_i}/\beta_0}c_i(\mu_0),
\end{align}
where $\mathcal{L} =\alpha _s(\mu_k) /\alpha_s(\mu_0) , \beta _0=11-2/3{n_f}$, and the one-loop anomalous dimensions $\gamma_{c_i}$ can be seen in Ref.~\cite{Fu:2020uzy}.

Next, in order to determine the continuum threshold and Borel parameters for the $D_s^+ \to \pi^0$ TFFs, one can follow the four criteria: (a) The continuum contributions are less than 30\% of the total results; (b) The contribution from the twist-4 LCDAs do not exceed 5\%; (c) We reuire the variations of the TFF within the Borel window be less than 10\%; (d) The continuum threshold $s_0$ should be closer to the squared mass of the first excited state of $D_s$-meson. Based on the fourth term of the criteria, we take $s_0$ to be close to the squared mass of the excited state of  $D_s$-meson $D_{s0}$(2590), {\it i.e.} $s_0 = 6.7(0.2)$ GeV$^2$. The reasonable Borel window is found to be $ M^2=25(2)\mathrm{GeV}^2$.

\begin{figure*}[t]
\begin{center}
\includegraphics[width=0.42\textwidth]{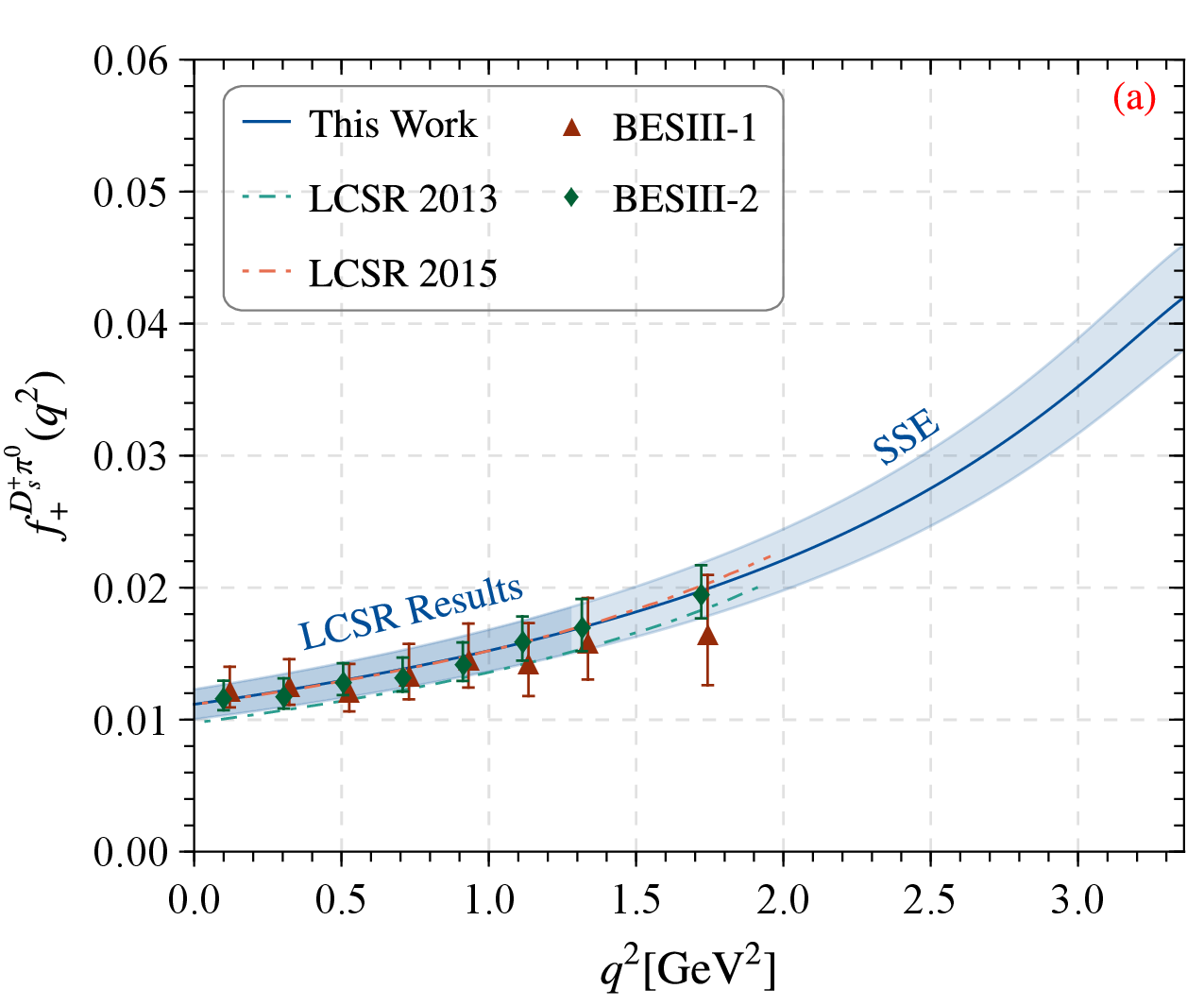}~~\includegraphics[width=0.425\textwidth]{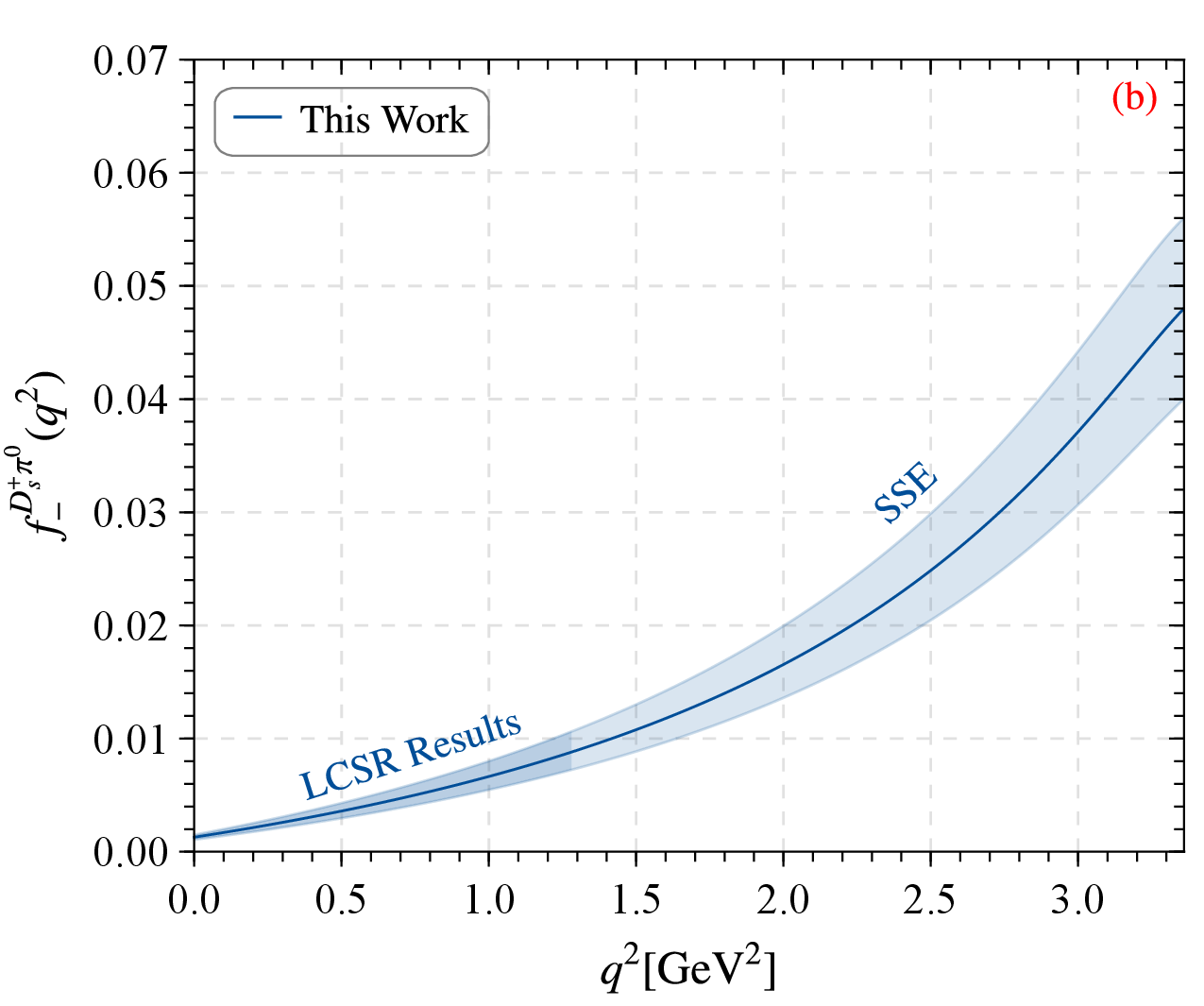}
\end{center}
\caption{The extrapolated LCSR predictions TFFs $f_\pm^{D_s^+\pi^0}(q^2)$ for the semileptonic decay $D_s^+\to\pi^0\ell^+\nu_\ell$ within uncertainties. Where, the darker and lighter bands stand for the LCSR results and SSE of our predictions. Results of the LCSR 2013~\cite{Offen:2013nma}, LCSR 2015~\cite{Duplancic:2015zna} and two set of BESIII collaboration~\cite{BESIII:2019qci} are presented as the comparison.}
\label{Fig:2}
\end{figure*}

Based on the parameters that have been determined, we can get the $D_s^+ \to \pi^0$ TFF at large recoil point $f_\pm^{D_s^+\pi^0}(0)$ with respect to each different input parameters, which can be arranged as follows,
\begin{align}
f_+^{D_s^+\pi^0}\left( 0 \right)&=0.0113+\left( _{-0.0008}^{+0.0008} \right) _{\delta}+\left( _{-0.0001}^{+0.0002} \right) _{s_0}\nonumber \\
	&+\left( _{-0.0000}^{+0.0001} \right) _{M^2}+\left( _{-0.0007}^{+0.0009} \right) _{m_c,f_{D_s}}+\left( _{-0.0002}^{+0.0003} \right) _{f_{\eta}}\nonumber \\
	&+\left( _{-0.0000}^{+0.0001} \right) _{a_{2;\eta}^2}+\left( _{-0.0000}^{+0.0000} \right) _{a_{4;\eta}^2}+\left( _{-0.0001}^{+0.0000} \right) _{a_{6;\eta}^2}\nonumber \\
	&=0.0113_{-0.0019}^{+0.0024},
\\
f_-^{D_s^+\pi^0}\left( 0 \right)&=0.0020+\left( _{-0.0002}^{+0.0001} \right) _{\delta}+\left( _{-0.0001}^{+0.0001} \right) _{s_0}\nonumber \\
	&+\left( _{-0.0000}^{+0.0000} \right) _{M^2}+\left( _{-0.0003}^{+0.0003} \right) _{m_c,f_{D_s}}+\left( _{-0.0001}^{+0.0001} \right) _{f_{\eta}}\nonumber \\
	&+\left( _{-0.0001}^{+0.0001} \right) _{a_{2;\eta}^2}+\left( _{-0.0000}^{+0.0000} \right) _{a_{4;\eta}^2}+\left( _{-0.0001}^{+0.0001} \right) _{a_{6;\eta}^2}\nonumber \\
	&=0.0020_{-0.0009}^{+0.0008}.
\end{align}

\begin{figure}[b]
\begin{center}
\includegraphics[width=0.425\textwidth]{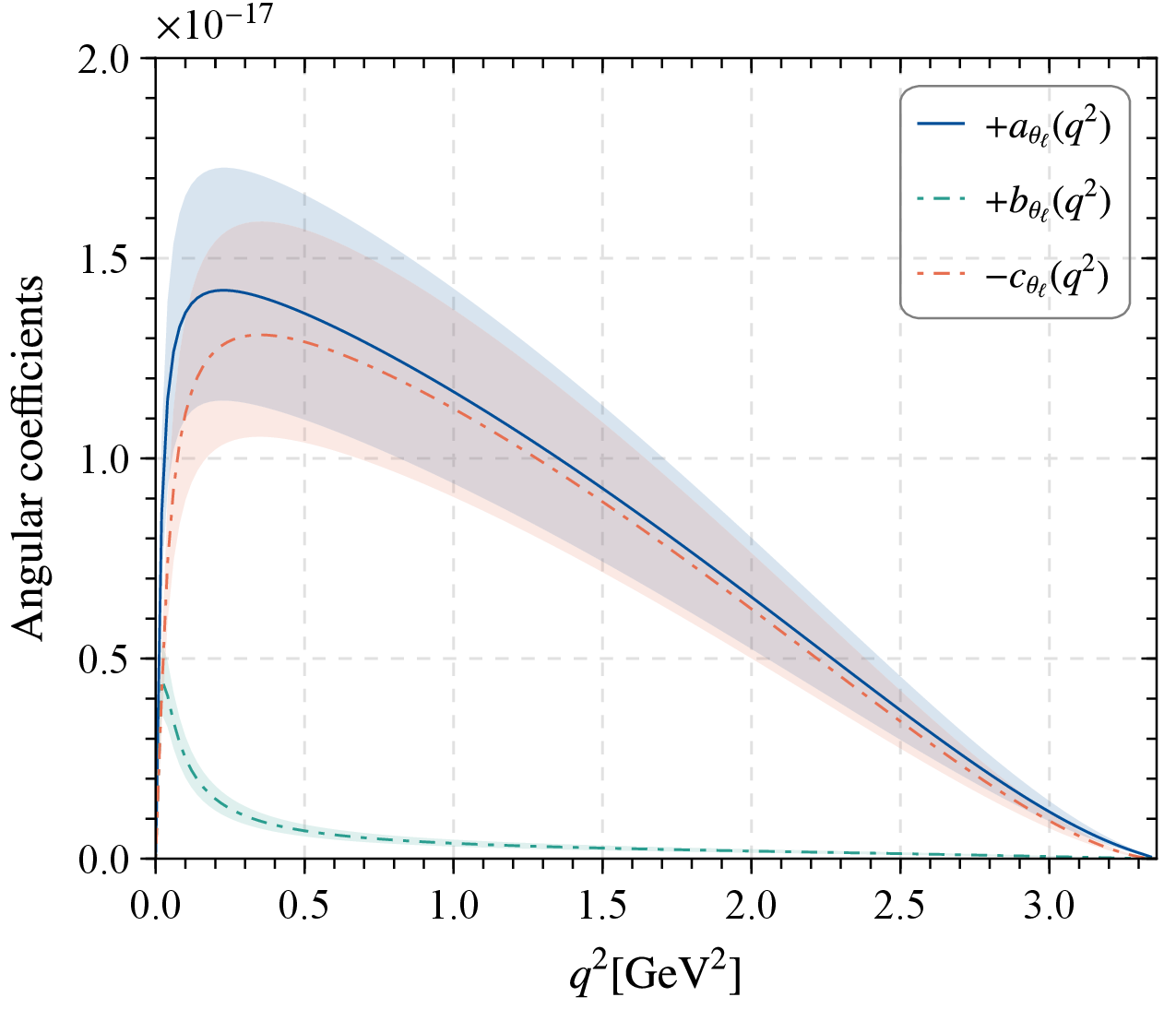}
\end{center}
\caption{The distribution of three $q^2$-dependent angular coefficient functions $a_{\theta_\ell}(q^2)$, $b_{\theta_\ell}(q^2)$ and $c_{\theta_\ell}(q^2)$ (in unit: $10^{-17}$), where the shaded bands stand for the uncertainties.}
\label{Fig:7}
\end{figure}

\begin{table}[t]
\footnotesize
\begin{center}
\caption{The fitting parameters for TFF $f_{+}^{D_s^+\pi^0}$ with central value (C), upper limits (U) and lower limits (L), which are represented as $f_{+;({\rm C,U,L})}^{D_s^+\pi^0}$ respectively.}
\label{Tab:SSEcoefficients}
\begin{tabular}{llll}
\hline
~~~~~~~~~~~~~~~~~~~~~~& $f_{+;{\rm C}}^{D_s^+\pi^0}$~~~~~~~~~~~~~~~~~~& $f_{+;{\rm U}}^{D_s^+\pi^0}$~~~~~~~~~~~~~~~~~~& $f_{+;{\rm L}}^{D_s^+\pi^0}$  \\  \hline
$\beta_0$   & $ 0.011$   & $-0.006$   & $-0.062$\\
$\beta_1$   & $ 0.012$   & $-0.007$  & $-0.070$\\
$\beta_2$   & $ 0.010$   & $-0.005$  & $-0.054$\\
$\Delta$    & $ 0.62\%$  & $ 0.67\%$ & $0.57\%$\\
\hline
\end{tabular}
\end{center}
\end{table}

The physical allowable range for the TFFs are $m_\ell^2 \leqslant q^2 \leqslant (m_{D_s^+}-m_{\pi^0})^2 \approx 3.36$ GeV$^2$. Theoretically, the LCSRs approach for $D_s^+ \to \pi^0\ell^+\nu_\ell$ TFFs are applicable in low and intermediate $q^2$-regions, {\it i.e.} $q^2 \in [0,1.3]$ GeV$^2$ of $\pi^0$-meson. One can extrapolate the TFFs in all physically allowable $q^2$-region via  $z(q^2,t)$ converging the simplified series expansion (SSE), {\it i.e.} the TFFs are expanded as~\cite{Bharucha:2015bzk}:
\begin{eqnarray}
f_{\pm}^{D_s^+\pi^0}(q^2) =\frac{1}{1-q^2/m_{D_s^2}}\sum_{k=0,1,2}{\beta _kz^k( q^2,t_0 )}
\end{eqnarray}
where $\beta_k$ are real coefficients and $z(q^2,t)$ is the function,
\begin{eqnarray}
z^k( q^2,t_0 ) =\frac{\sqrt{t_+-q^2}-\sqrt{t_+-t_0}}{\sqrt{t_+-q^2}+\sqrt{t_+-t_0}},
\end{eqnarray}
with $t_{\pm} = (m_{D_s^+} \pm m_{\pi^0})^2$ and $t_0=t_{\pm}(1-\sqrt{1-t_-/t_+})$. The SSE method possesses superior merit, which keeps the analytic structure correct in the complex plane and ensures the appropriate scaling, $f_{\pm}^{D_s^+\pi^0}(q^2)\sim 1/q^2$ at large $q^2$. And the quality of fit $\Delta$ is devoted to take stock of the resultant of extrapolation, which is defined as
\begin{eqnarray}
\Delta =\frac{\sum_t{| F_i(t) -F_{i}^{\mathrm{fit}}( t ) |}}{\sum_t{| F_i(t) |}}\times 100.
\end{eqnarray}

After making extrapolation for the TFFs $f_\pm^{D_s^+\pi^0}(q^2)$ to the whole physical $q^2$-region. Take  $f_+^{D_s^+\pi^0}(q^2)$ for instance, we listed the coefficients $\beta_{0,1,2}$ and $\Delta$ of its central value, upper limits and lower limits with symbols ``C'', ``U'' and ``L'' in Table~\ref{Tab:SSEcoefficients}, respectively. The quality of fit is lower than 1\%, which shows the higher agreement between SSE and LCSR results. Then, the behaviors of $D_s^+\to\pi^0$ TFFs in the whole physical region with respect to squared momentum transfer are given in Fig.~\ref{Fig:2}, where the darker and lighter bands stand for the LCSR results and SSE of our predictions. As a comparison, we also present the predictions from theoretical and experimental groups, such as the LCSR 2013~\cite{Offen:2013nma}, the LCSR 2015~\cite{Duplancic:2015zna}, and the two set of BESIII collaboration~\cite{BESIII:2019qci}. Here, we have a notation that the theoretical and experimental results are coming from the relationship Eq.~\eqref{Eq::14} with the help of $D_s^+\to\eta$ TFFs. The type-1 set of BESIII result stands for $\eta\to\gamma\gamma$ channel and type-2 is $\eta\to\pi^0\pi^+\pi^-$ channel. The curves show that our results are in good agreement with other theoretical and experimental predictions within uncertainties. Furthermore, we present the behaviors of the three angular coefficients functions $a_{\theta_\ell}(q^2)$, $b_{\theta_\ell}(q^2)$ and $c_{\theta_\ell}(q^2)$ uncertainties with unit $10^{-17}$-order level in Fig.~\ref{Fig:7}. The negative of $c_{\theta_\ell}$ is given for convenience to compare the three angular coefficients. As can be seen from the figure, the absolute values of $a_{\theta_\ell}(q^2)$ and $c_{\theta_\ell}(q^2)$ are very closer with uncertainties, and value for $b_{\theta_\ell}(q^2)$ is smaller than that of $a_{\theta_\ell}(q^2)$ and $-c_{\theta_\ell}(q^2)$.

\begin{figure}[t]
\begin{center}
\includegraphics[width=0.42\textwidth]{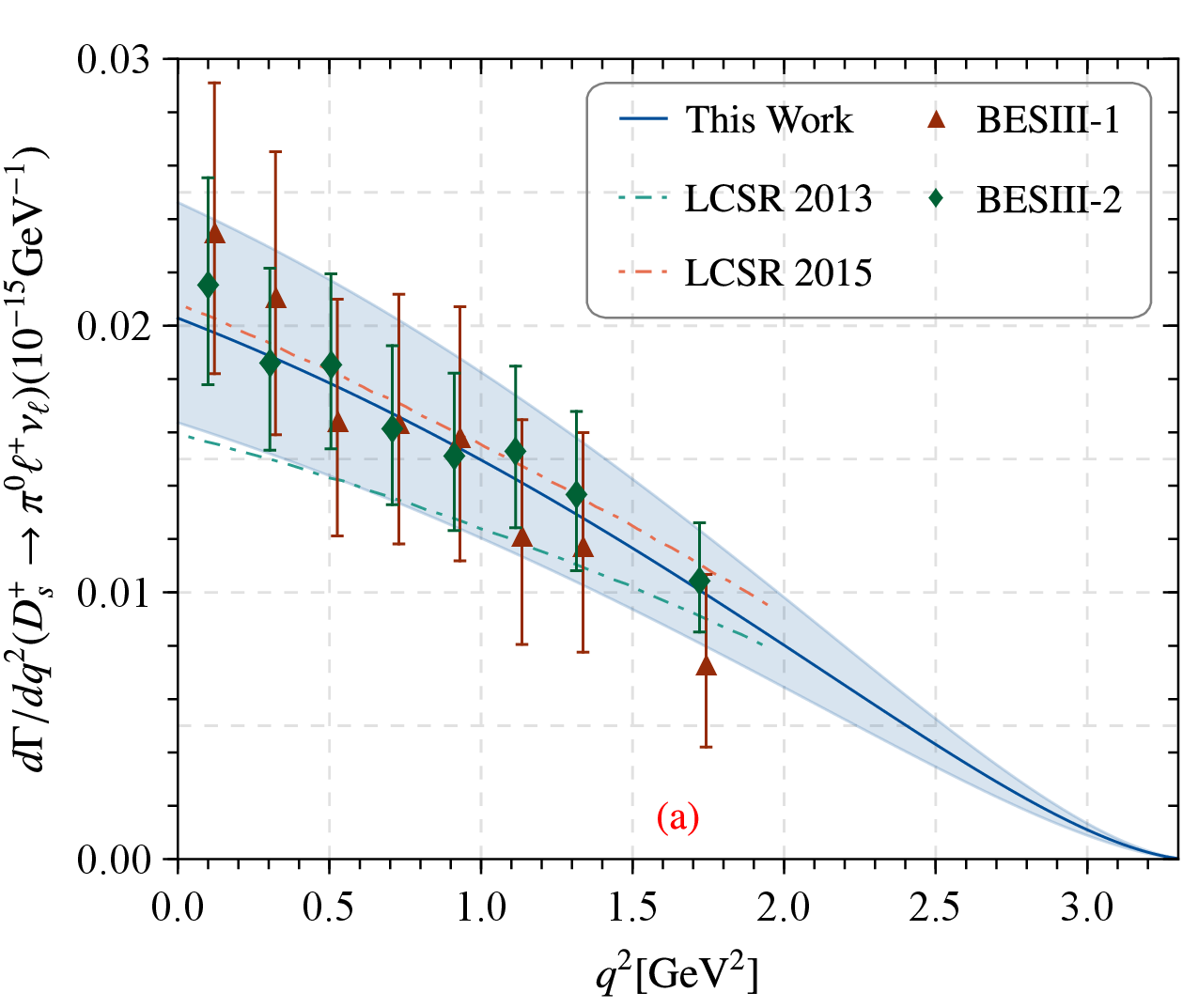}
\includegraphics[width=0.42\textwidth]{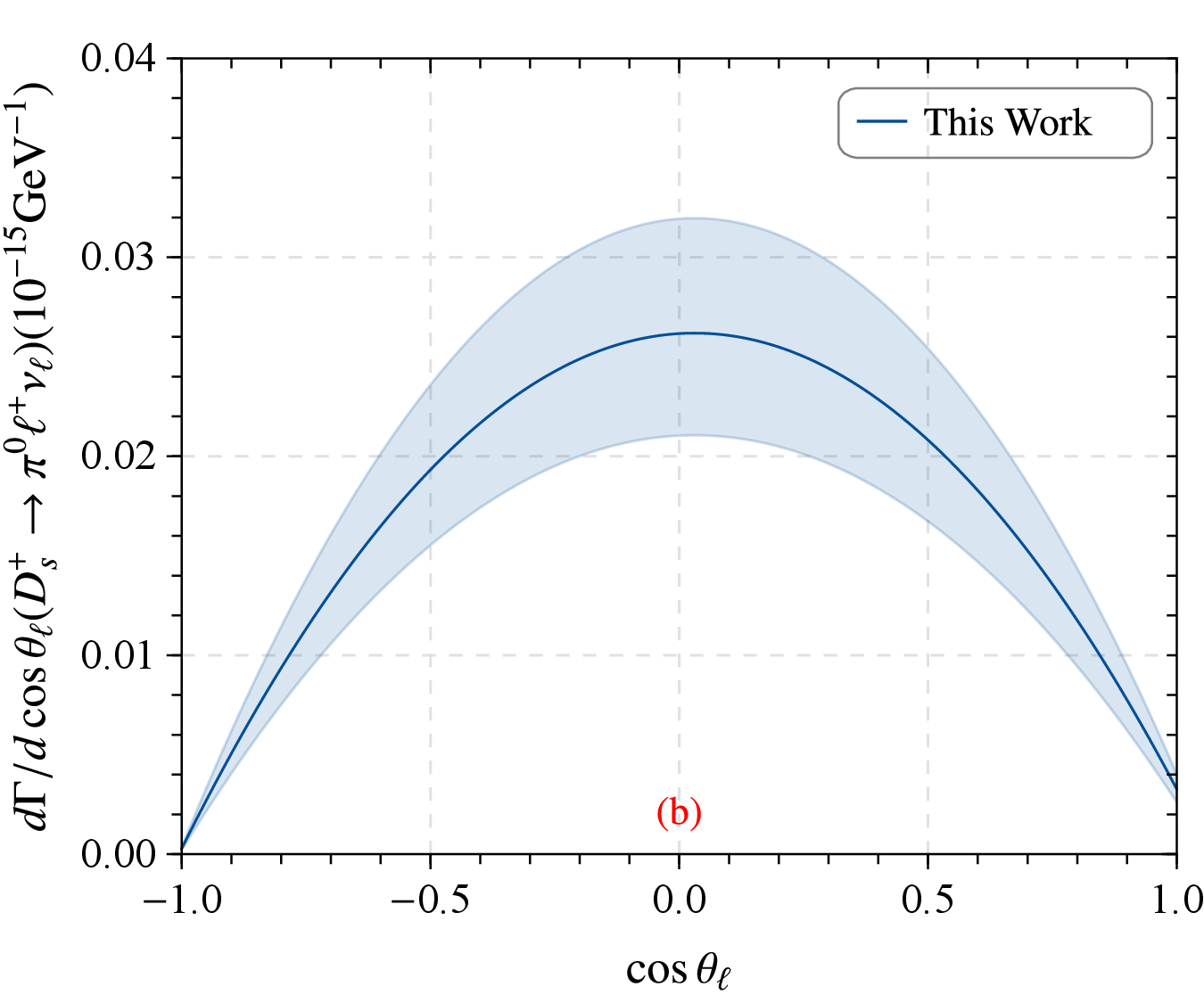}
\end{center}
\caption{Differential decay width for the $D_s^+ \to \pi^0 \ell^+ \nu_\ell$ versus $q^2$ and $\cos\theta_\ell$, with shaded bands corresponds to uncertainties. Meanwhile, results from LCSR 2013~\cite{Offen:2013nma}, LCSR 2015~\cite{Duplancic:2015zna} and two set of BESIII~\cite{BESIII:2019qci} are presented as the comparison.}
\label{Fig:4}
\end{figure}

For the next stage, we comment on some phenomenological results for semileptonic decay $D_s^+\to\pi^0 \ell^+ \nu_\ell$, {\it i.e.} the decay width, branching fraction, lepton-flavor universality, and other observables. In which the CKM matrix element $|V_{cs}|$ is required. Here, we mainly take the average value of the leptonic and semileptonic decay processes for $c\to s$, which comes from PDG~\cite{ParticleDataGroup:2020ssz}, i.e $|V_{cs}|=0.987\pm 0.011$. With the resultant TFFs, we present the $D_s^+\to \pi^0\ell^+\nu_\ell$ full differential decay width with respect to the two kinematic variables: squared momentum transfer $q^2$ and cosine angle $\cos\theta_\ell$ in Fig.~\ref{Fig:4}, which has the following notations:
\begin{itemize}
  \item As a comparison, we present the predictions from the LCSR in 2013~\cite{Offen:2013nma} and 2015~\cite{Duplancic:2015zna},  the BESIII~\cite{BESIII:2019qci} in Fig.~\ref{Fig:4}(a), which are also obtained from the $D_s^+ \to \eta \ell^+\nu_\ell$ by using the expression Eq.~\eqref{Eq::14}.
  \item In Fig.~\ref{Fig:4}(a), our predictions have agreement with other LCSR results and BESIII data within errors in the region $0 \leqslant q^2 \leqslant 1.95~{\rm GeV^2}$. And the curves of our predictions tend to zero when the squared momentum transfer leans towards the small recoil region.
  \item In Fig.~\ref{Fig:4}(b), we present the angular distribution $d\Gamma(D_s^+ \to \pi^0 \ell^+ \nu_\ell)/d\cos\theta_\ell$ in the region of $-1\leqslant \cos\theta_\ell\leqslant 1$, and the curve is asymmetry.
  \item The uncertainties of our predictions are mainly coming from each input theoretical parameters.
\end{itemize}

\begin{table}[b]
\footnotesize
\begin{center}
\caption{The  $D_s^+\to\pi^0\ell^+\nu_\ell$ with $\ell = (e , \mu)$ total branching fractions (in unit: $10^{-5}$) within uncertainties. Meanwhile, the neutral meson mixing effect (NMME) from Li and Yang~\cite{Li:2020ylu}, and also the BESIII collaboration upper limits~\cite{BESIII:2022jcm} are presented as a comparison.}
\label{tab:2}
\begin{tabular}{lll}
\hline
~~~~~~~~~~~~~~~~~~~~~~~~~&$\mathcal{B} (D_s^+\to\pi^0e^+\nu_e) $~~~~~~~~&$\mathcal{B}( D_s^+\to\pi^0\mu^+\nu _\mu )$\\ \hline
This work                                  &$2.60_{-0.51}^{+0.57}$  & $2.58_{-0.51}^{+0.56}$\\
NMME~\cite{Li:2020ylu}           & $2.65\pm0.38$            & $-$\\
BESIII~\cite{BESIII:2022jcm}   & $< 6.4$     & $-$ \\
\hline
\end{tabular}
\end{center}
\end{table}

After integrating over the whole $q^2$-region, {\it i.e.} $m_\ell^2 \leqslant q^2 \leqslant (m_{D_s}-m_{\pi^0})^2 \approx 3.36$ GeV$^2$ for differential decay widths, we obtain the total decay widths for $D_s^+ \to \pi^0 \ell^+ \nu_\ell$ with two different channels
\begin{align}
&\Gamma \left( D_s^+\rightarrow \pi ^0e^+\nu_e \right) =0.0339_{-0.0066}^{+0.0074}\times 10^{-15} ~{\mathrm{GeV}},
\label{Eq:32}\\
&\Gamma \left( D_s^+\rightarrow \pi ^0\mu ^+\nu _{\mu} \right) =0.0337_{-0.0066}^{+0.0074}\times 10^{-15} ~{\mathrm{GeV}},
\label{Eq:33}
\end{align}
which have slight different with each other. Furthermore, after using the lifetime of initial state $D_s^+$-meson, {\it i.e.} $\tau _{D_s^+}=( 0.504\pm 0.007 )$ ps~\cite{ParticleDataGroup:2020ssz}, we can get the branching fraction for the semileptonic decay channels $D_s^+ \to \pi^0 \ell^+ \nu_\ell$ with $\ell=(e,\mu)$. The results are presented in Table~\ref{tab:2}. The neutral meson mixing effect (NMME) from Li and Yang~\cite{Li:2020ylu}, and also the BESIII collaboration upper limits~\cite{BESIII:2022jcm} are presented as a comparison. The result for $D_s^+ \to \pi^0 e^+ \nu_e$ channel shows that our prediction have agreement with Li and Yang, which both in the reasonable region predicted by BESIII collaboration. We present the $D_s^+ \to \pi^0 \mu^+ \nu_\mu$ simultaneously.

\begin{figure}[t]
\begin{center}
\setstretch{0.9}
\includegraphics[width=0.385\textwidth]{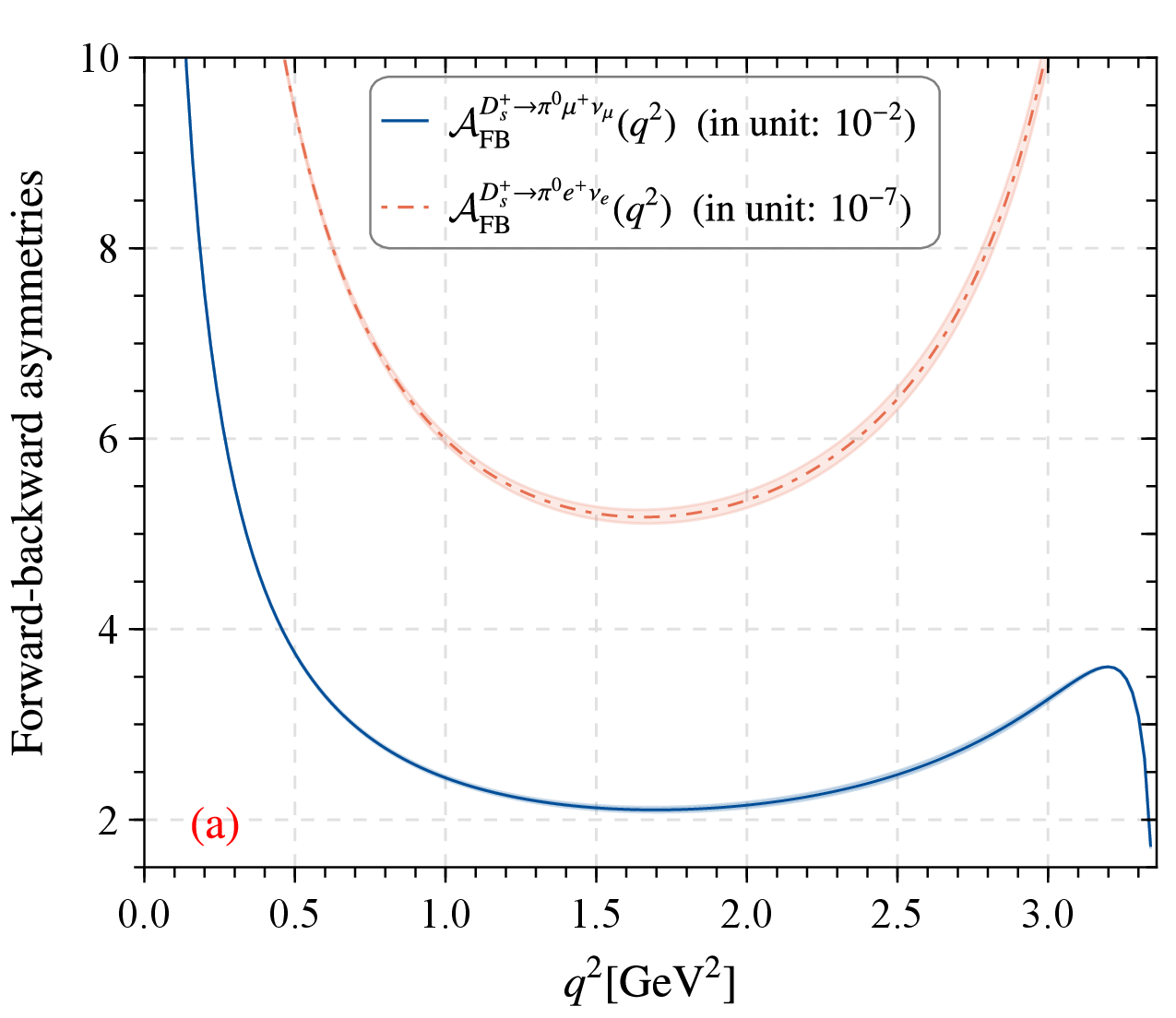}
\includegraphics[width=0.385\textwidth]{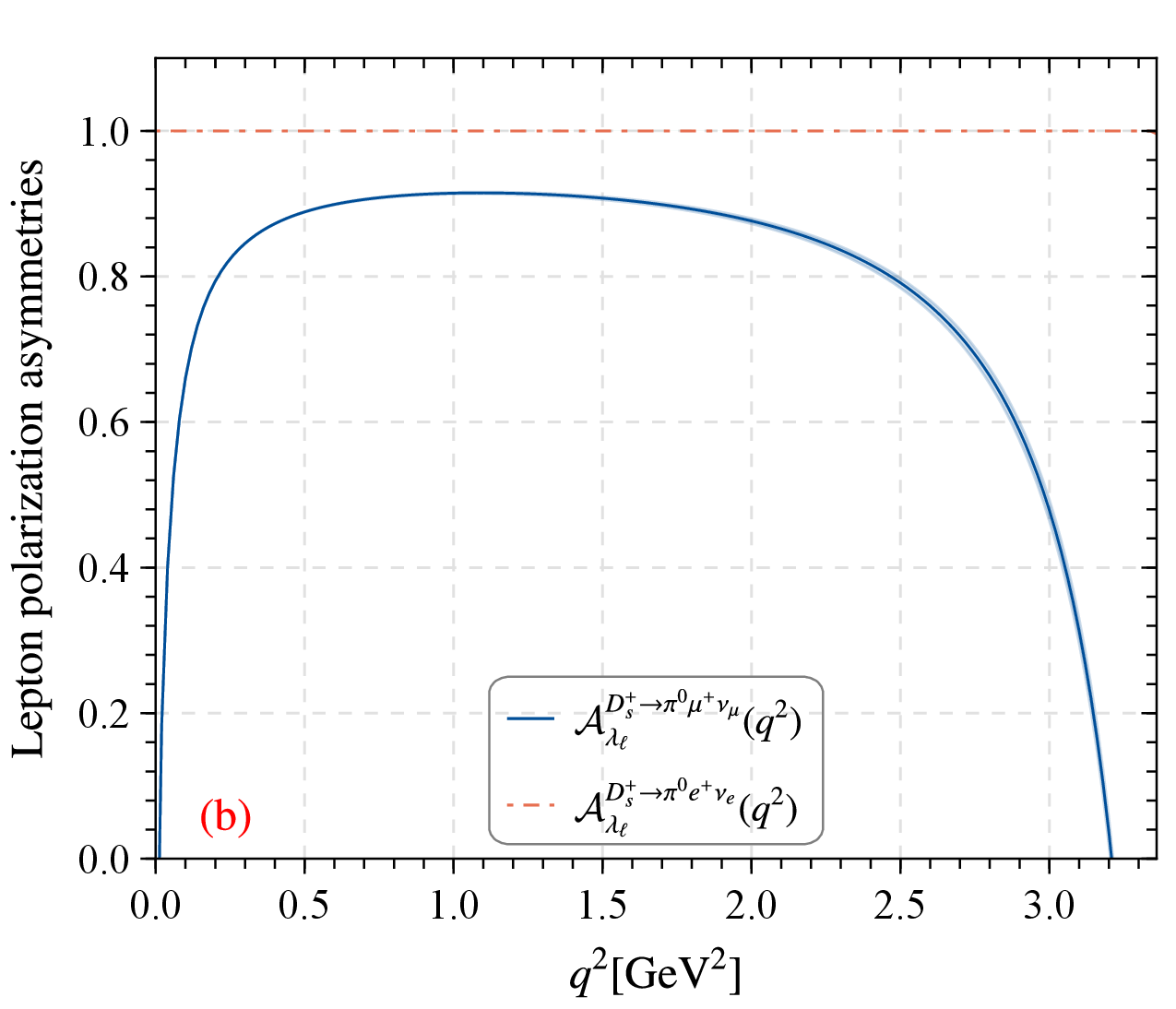}
\includegraphics[width=0.385\textwidth]{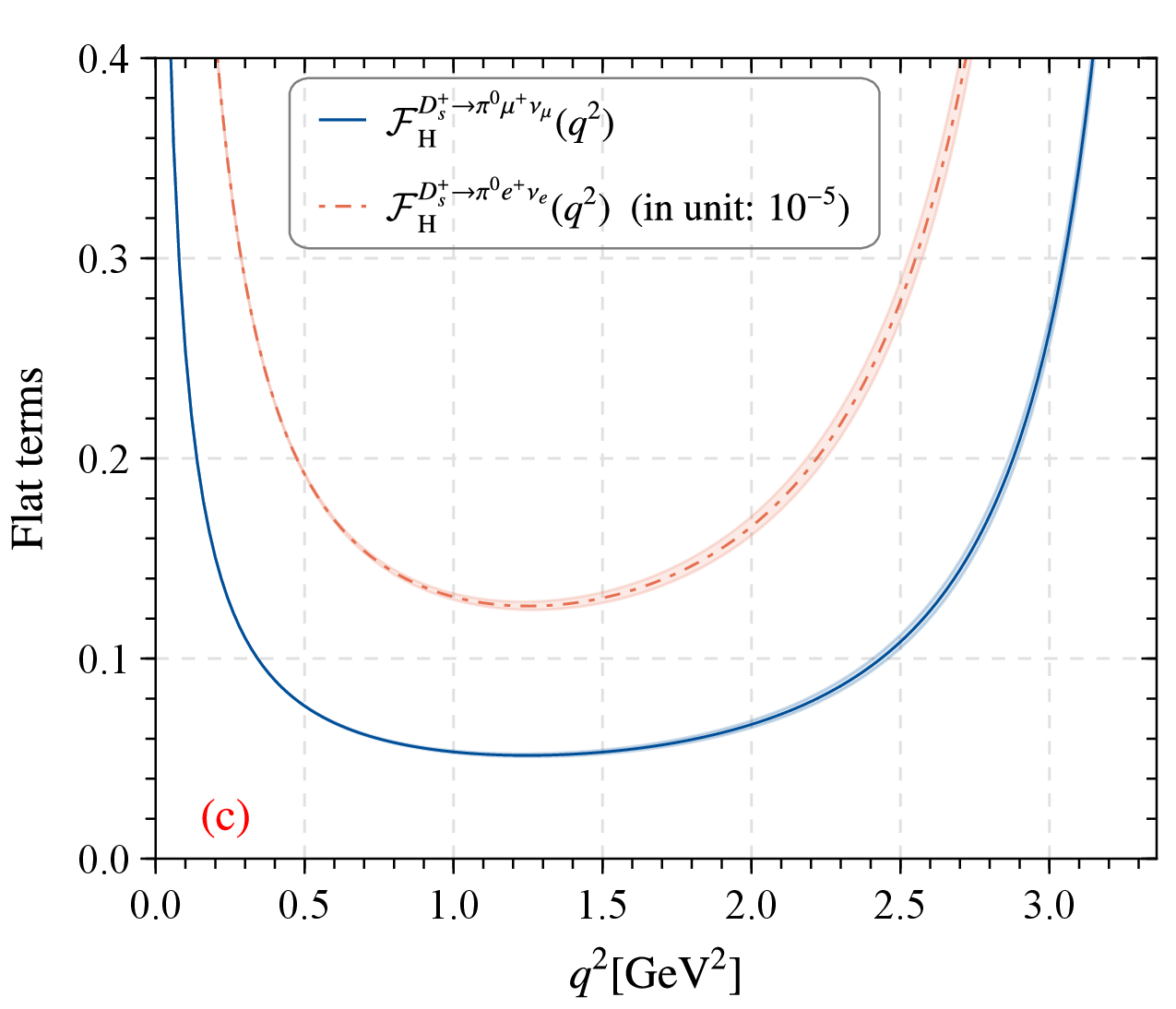}
\end{center}
\caption{Theory prediction for the three various items angular observables $\mathcal{A}_{\mathrm{FB}}^{D_s^+\to\pi^0\ell^+\nu_\ell}(q^2)$, $\mathcal{F}_{\mathrm{H}}^{D_s^+\to\pi^0\ell^+\nu_\ell}(q^2)$, and $\mathcal{A}_{\lambda_\ell}^{D_s^+\to\pi^0\ell^+\nu_\ell}(q^2)$ with $\ell = (e,\mu)$. In which, shaded bands are the uncertainties.}
\label{Fig:6}
\end{figure}

As a further step, the three differential distribution of angle observables of the semileptonic decay $D_s^+\to\pi^0\ell^+\nu_\ell$ with $\ell = (e,\mu)$, {\it i.e.} the forward-backward asymmetries $\mathcal{A}_{\mathrm{FB}}^{D_s^+\to\pi^0\ell^+\nu_\ell}(q^2)$, the $q^2$-differential flat terms $\mathcal{F}_{\mathrm{H}}^{D_s^+\to\pi^0\ell^+\nu_\ell}(q^2)$, and lepton polarization asymmetries $\mathcal{A}_{\lambda_\ell}^{D_s^+\to\pi^0\ell^+\nu_\ell}(q^2)$ are presented in Fig.~\ref{Fig:6}, which shows that
\begin{itemize}
  \item Their center values are nearly equal to the upper/lower limits in the region $0\leqslant q^2\leqslant 2.0~{\rm GeV^2}$, and have slight difference in  $2.0~{\rm GeV^2}< q^2\leqslant 3.36~{\rm GeV^2}$. This is agreement with the $B\to\pi(K)\ell\nu_\ell$ cases~\cite{Cui:2022zwm}
  \item Due to the massless of electron, the distribution of lepton polarization asymmetry within uncertainties equal to 1.
  \item The electron channel is about $10^{-5}$-order lower than the muon channel for the forward-backward asymmetries and flat terms.
\end{itemize}
The integrated results of the three angular observables are
\begin{align}
&\mathcal{A}_{\mathrm{FB}}^{D_s^+\to\pi^0\mu^+\nu_\mu} = 1.22_{-0.01}^{+0.01}\times 10^{-1},                         \\
&\mathcal{A}_{\mathrm{FB}}^{D_s^+\to\pi^0e^+\nu_e} = 7.49_{-0.02}^{+0.02}\times 10^{-6},               \\
&\mathcal{F}_{\mathrm{H}}^{D_s^+\to\pi^0\mu^+\nu_\mu} = 0.48_{-0.01}^{+0.01},                                 \\
&\mathcal{F}_{\mathrm{H}}^{D_s^+\to\pi^0e^+\nu_e} = 1.77_{-0.14}^{+0.13}\times 10 ^{-5},                       \\
&\mathcal{A}_{\lambda_\ell}^{D_s^+\to\pi^0\mu^+\nu_\mu} = 2.49_{-0.02}^{+0.01},                                \\
&\mathcal{A}_{\lambda_\ell}^{D_s^+\to\pi^0e^+\nu_e} = 3.36_{-0.00}^{+0.00}.
\end{align}
Finally, the specific value of the ratio for the different decay channels ${\cal R} _{\pi ^0/\eta}^{\ell}$ is presented as follows:
\begin{eqnarray}
{\cal R} _{\pi ^0/\eta}^{e} &=& \frac{\mathcal{B} (D_s^+\to \pi ^0e^+\nu_e )}{\mathcal{B} ( D_s^+\to \eta e^+\nu_e)}
\nonumber  \\
&=& 1.108_{-0.071}^{+0.039}\times 10^{-3},
\end{eqnarray}
where the branching fraction is ${\mathcal{B} \left( D_s^+\to \eta e^+\nu_e \right)}=2.346_{-0.331}^{+0.418}\times10^{-2}$, which are taken from our previous work~\cite{Hu:2021zmy}. This can be considered as a good test for the correctness of the considered $D_s^+-$meson internal structure, and also the mixing angle between $\pi^0$ and $\eta$ states.

\section{Summary}\label{Sec:4}
In order to have a deeper insight into heavy-to-light decay, we carry out the study of semileptonic decay $D_s^+ \to \pi^0 e^+ \nu_e$ in this paper. Firstly, the mechanism of neutral meson mixing effect is introduced briefly, the $D_s^+\to \pi^0$ TFFs $f_\pm^{D_s^+\to \pi^0}(q^2)$ are investigated within LCSR approach up to NLO correction, the $\eta$-meson with $s\bar s$ component twist-2 LCDA is researched by QCD sum rule under background field theory up to full dimension-six accuracy. Secondly, we make the extrapolation for $f_\pm^{D_s^+\to \pi^0}(q^2)$ to the whole $q^2$-region $m_\ell^2\leqslant q^2 \leqslant (m_{D_s^+}-m_{\pi^0})^2$ by using SSE, and make a comparison with BESIII and other theoretical group. The behaviors of three TFFs related angular coefficients functions $a_{\theta_\ell}$, $b_{\theta_\ell}$, $c_{\theta_\ell}$ are presented.

Then, the differential decay width for $D_s^+ \to \pi^0 \ell^+ \nu_\ell$ versus $q^2$ and $\cos\theta_\ell$ within uncertainties are presented in Fig.~\ref{Fig:4}. Our result shows well agreement with BESIII collaboration and other LCSR predictions. The total decay width results are given in Eqs.~\eqref{Eq:32} and \eqref{Eq:33}. Furthermore, after considering the lifetime of initial state, we obtained the branching fraction for the semileptonic decay channels $D_s^+ \to \pi^0 \ell^+ \nu_\ell$ with $\ell=(e,\mu)$. The results are presented in Table~\ref{tab:2}. Our prediction have agreement with Li and Yang, which both in the reasonable region predicted by BESIII collaboration. Finally, we make analysis about the forward-backward asymmetries, the $q^2$-differential flat terms, lepton polarization asymmetries, and also the ratio for different decay channel ${\cal R} _{\pi ^0/\eta}^{e}=1.108_{-0.071}^{+0.039}\times 10^{-3}$.

\acknowledgments
We are grateful to the Prof. Maxim Yu.~Khlopov, Prof. Xian-Wei Kang and Referee for their valuable comments and suggestions. This work was supported in part by the National Natural Science Foundation of China under Grant No.12265010, No.12265009, the Project of Guizhou Provincial Department of Science and Technology under Grant No.ZK[2021]024 and No.ZK[2023]142, the Project of Guizhou Provincial Department of Education under Grant No.KY[2021]030.


\begin{thebibliography}{99}
\bibitem{Zweig:1964ruk}
    G.~Zweig,
    An SU(3) model for strong interaction symmetry and its breaking. Version 1,2,.
\end{thebibliography}
\end{document}